\pdfoutput=1

\documentclass[10pt,a4paper,twocolumn, twoside]{article}
\usepackage{graphicx}
\usepackage[T1]{fontenc}
\usepackage[latin1]{inputenc}
\usepackage{amsmath}
\usepackage{caption}
\usepackage{booktabs}
\usepackage{setspace}
\usepackage{etoolbox}

\AtBeginEnvironment{tabular}{\onehalfspacing}

\makeatletter
\renewcommand{\section}{\@startsection{section}{2}{0cm}{-\baselineskip}
{0,5\baselineskip}{\normalsize\bfseries}}
\renewcommand{\subsection}{\@startsection{subsection}{3}{0cm}{-\baselineskip}
{0,5\baselineskip}{\normalsize\slshape}}
\makeatother

\RequirePackage{mathptmx}      
\RequirePackage[numbers,sort&compress]{natbib}
\RequirePackage[colorlinks,citecolor=blue,urlcolor=blue,linkcolor=blue]{hyperref}
\RequirePackage{siunitx}
\newcommand*{\xe}[1]{\ensuremath{{}^{#1}\textrm{Xe}}}
\newcommand*{\kr}[1]{\ensuremath{{}^{#1}\textrm{Kr}}}
\newcommand*{\ar}[1]{\ensuremath{{}^{#1}\textrm{Ar}}}
\newcommand*{\xepip}{xenon sampling device}
\newcommand*{\perc}[1]{\SI{#1}{\percent}}
\newcommand*{\degc}[1]{\SI{#1}{\degreeCelsius}}
\newcommand*{\units}[1]{\ensuremath{\, \textrm{#1}}}

\DeclareSIUnit[]\ppb{ppb}
\DeclareSIUnit[]\ppt{ppt}
\DeclareSIUnit[]\ppq{ppq}
\DeclareSIUnit[]\ccm{\cubic\cm}
\DeclareSIUnit[]\mbar{\milli\bar}

\begin{document}

\title{Krypton assay in xenon at the ppq level
using a gas chromatographic system and mass
spectrometer}

\author{Sebastian Lindemann and Hardy Simgen}

\date{\small \it
Max-Planck-Institut f\"ur Kernphysik, Saupfercheckweg 1, D-69117 Heidelberg, Germany \\
\vspace{0.3cm}
Email-addresses: \\ {\tt Sebastian.Lindemann@mpi-hd.mpg.de\\Hardy.Simgen@mpi-hd.mpg.de} \\
\vspace{0.3cm}
{\it (Published in \href{http://dx.doi.org/10.1140/epjc/s10052-014-2746-1}{Eur. Phys. J. C (2014) 74:2746})}
}

\twocolumn[
\begin{@twocolumnfalse}
\maketitle

\begin{abstract}
\noindent	We have developed a new method to measure krypton traces in xenon at
	unprecedented low concentrations. This is a mandatory task for many
	near-future low-background particle physics detectors.  Our system
	separates krypton from xenon using cryogenic gas chromatography. The
	amount of krypton is then quantified using a mass spectrometer.  We
	demonstrate that the system has achieved a detection limit of 8 ppq
	(parts per quadrillion) and present results of distilled xenon with
	krypton concentrations below 1 ppt.
\end{abstract}
\end{@twocolumnfalse}
]

\section{\label{sec:intro}Introduction}

Low-background particle physics experiments rely on liquid xenon (LXe)
detectors in the fields of direct dark matter detection, neutrinoless
double beta decay and solar neutrino research
\cite{Aprile:2011dd,2012JInst...7.5010A,2013NIMPA.716...78A}.  LXe is used
as a target material due to its high mass, good self-shielding and
scintillation properties. Xenon is commercially produced by extraction from
the atmosphere. The relative abundances of krypton and xenon in ambient air
are 1.14 parts per million by volume (ppmv) and 0.09 ppmv, respectively
\cite{DIN_1871:1999-05}. The separation from atmospheric constituents in
this extraction process has limited efficiency resulting in traces of
krypton at the level of parts per million (ppm) and, for high purity xenon,
parts per billion (ppb).

The inert gas krypton with its unstable isotope \kr{85} is one of the most
serious internal background sources for LXe detectors used in
low-background experiments. It is homogeneously distributed in the xenon
and cannot be discriminated by
shielding or fiducial volume cuts.  \kr{85} has a half life of 10.8 years
and is an almost pure $\beta^{-}$ emitter (\perc{99.56} branching
ratio) with end-point energy of 687 keV \cite{Recommended_Data_LNHB}. The
activity of the man-made isotope \kr{85} in the atmosphere, produced in
sizable quantities by nuclear fission and released by nuclear-fuel
reprocessing plants and nuclear weapon tests, has been steadily increasing
over time. The present-day activity concentration is approximately
\SI{1.4}{\becquerel\per\cubic\meter} \cite{Loosli..2000,
bieringer-bfs..2009}, corresponding to a relative isotopic abundance of
\kr{85}/\kr{\textrm{nat}} of \SI{2e-11}{\mole\per\mole}
\cite{2003GeoRL..30tHLS4D}.

Purification techniques are established to remove krypton from xenon to the
parts per trillion (ppt) level \cite{2012JInst...7.5010A, 2009APh....31..290A}.
Techniques for the quantification of such ultra-low levels of contamination so
far only exist until the few hundreds ppq level
\cite{2009APh....31..290A,Dobi20111,2013RScI...84i3105A}.  However, krypton/xenon
concentrations below 100 parts per quadrillion (ppq) are needed for near future
experiments \cite{Aprile:2012zx}. 

In this work we present a measurement technique with an extrapolated
detection limit of \SI{8}{\ppq}, almost two orders of magnitude
lower than what has been previously achieved. Section~\ref{sec:exp_setup} describes
the measuring system based on gas chromatographic separation of the bulk
xenon from the krypton traces and quantification by means of a sector field
mass spectrometer. In Sect.~\ref{sec:calibration}, the data
analysis is described, the system acceptance for krypton is characterized,
a detailed study of the experimental uncertainties is presented and the system's
detection limit is given. Finally, in Sect.~\ref{sec:results} the results
of seven different samples that prove the claimed performance are shown and
the assay of xenon with a krypton concentration below the ppt level is
presented.

\section{Experimental setup}\label{sec:exp_setup}
The mass spectroscopic technique is used to quantify the abundance of
natural krypton in given xenon gas batches down to the ppq regime. The
setup can be separated into four parts: sample preparation, ion source,
mass analyzer and detector. The ion source, mass analyzer and detector belong
to a customized sector field mass spectrometer (Vacuum
Generators model VG 3600) that is described in Sect.~\ref{sec:ms}. The sample
preparation, based upon cryogenic gas chromatography, is the essential new
development and will be discussed in Sect.~\ref{sec:gc}.


\subsection{The mass spectrometer}\label{sec:ms}
\begin{figure}[b]
	\includegraphics[width=\columnwidth]{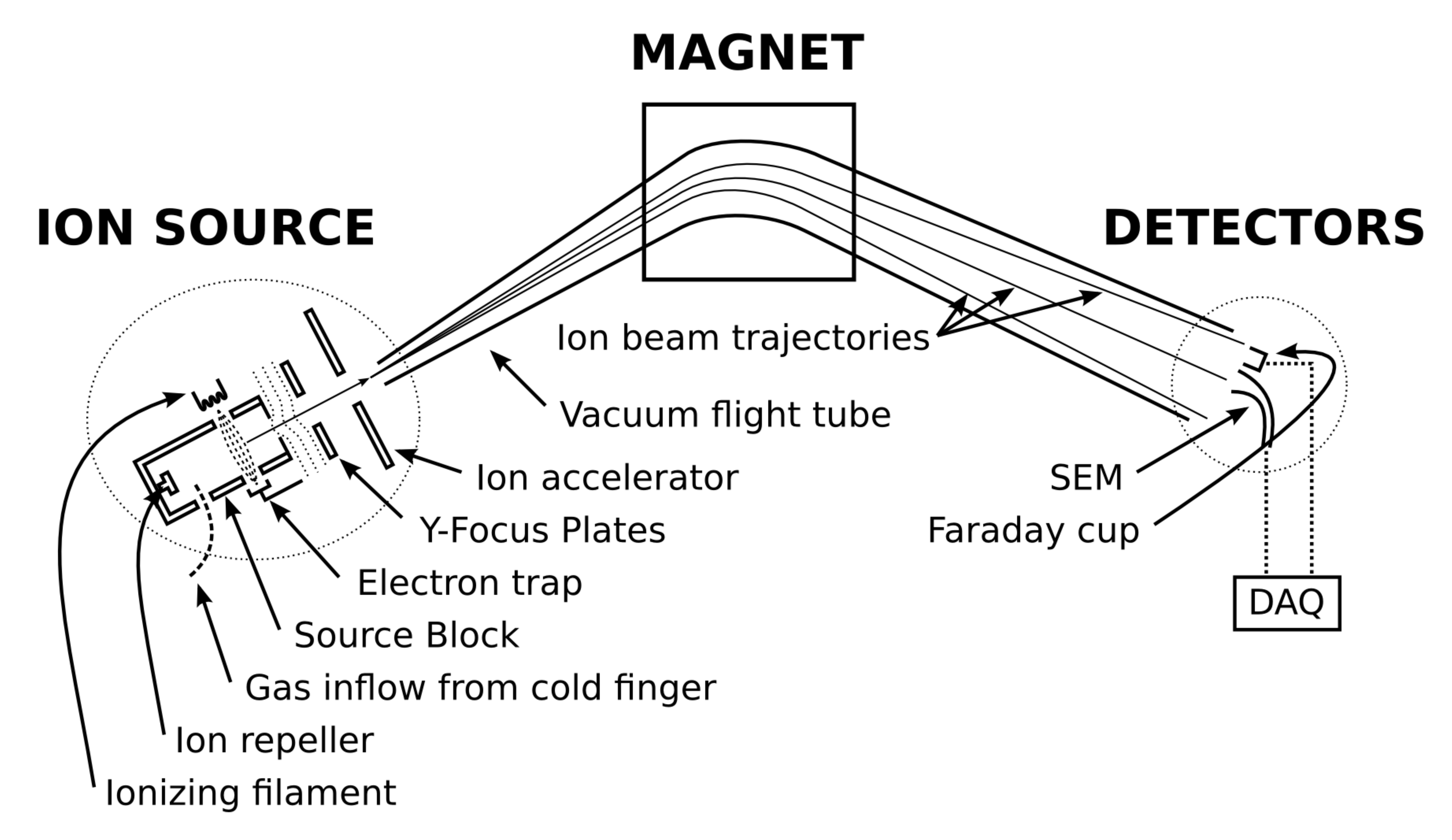}
	\caption{\label{fig:msonly}Scheme of the working principle of the mass spectrometer (VG
	3600) applied to quantify krypton traces down to \SI{e-13}{\cubic\cm}}
\end{figure}

The mass spectrometer is a customized version of a VG 3600
(Fig.~\ref{fig:msonly}) capable of quantifying an
amount of natural krypton of less than \SI{e-13}{\cubic\cm} STP\footnote{In
this work gas quantities will be always given at standard temperature and
pressure using the International Union of Pure and Applied Chemistry (IUPAC)
recommendation of $\degc{0}$ and \SI{100}{\kilo\pascal} \cite{iupac_golden_book}.}
\cite{Hopp2007635}. It is located at the Max-Planck-Institut f{\"u}r Kernphysik
in Heidelberg, Germany. 

The sample is collected by cryogenic pumping on a cold finger which
contains a small amount of activated carbon, which is installed between the
sample preparation part and ion source. When warming up the cold finger,
the gaseous sample distributes in the entire volume of ion source, mass
analyzer and detectors.  Ionization, acceleration and focusing take place
in the ion source of the mass spectrometer. Electrons are emitted from a
hot filament and ionize the sample gas. The ions are accelerated and
focused in electric fields. Their mass separation is achieved by a variable
magnetic dipole-field. The ions are detected by a continuous dynode
amplifier (SEM = secondary electron multiplier) counting events above a
predefined threshold optimized for its signal/background ratio.  In order
to achieve the high sensitivity goals batch sizes on the order of
\SI{1}{\ccm} xenon gas are needed. However, this amount of xenon gas would
result in a pressure much above the critical pressure of \SI{e-6}{\mbar}
\cite{Hoffmann2007} in the spectrometer. Thus, the krypton must be
separated from the bulk xenon before it is fed into the spectrometer. This
separation is achieved by the gas chromatography system.

\subsection{The gas chromatography system}\label{sec:gc}
\begin{figure}
  \includegraphics[width=0.95\columnwidth]{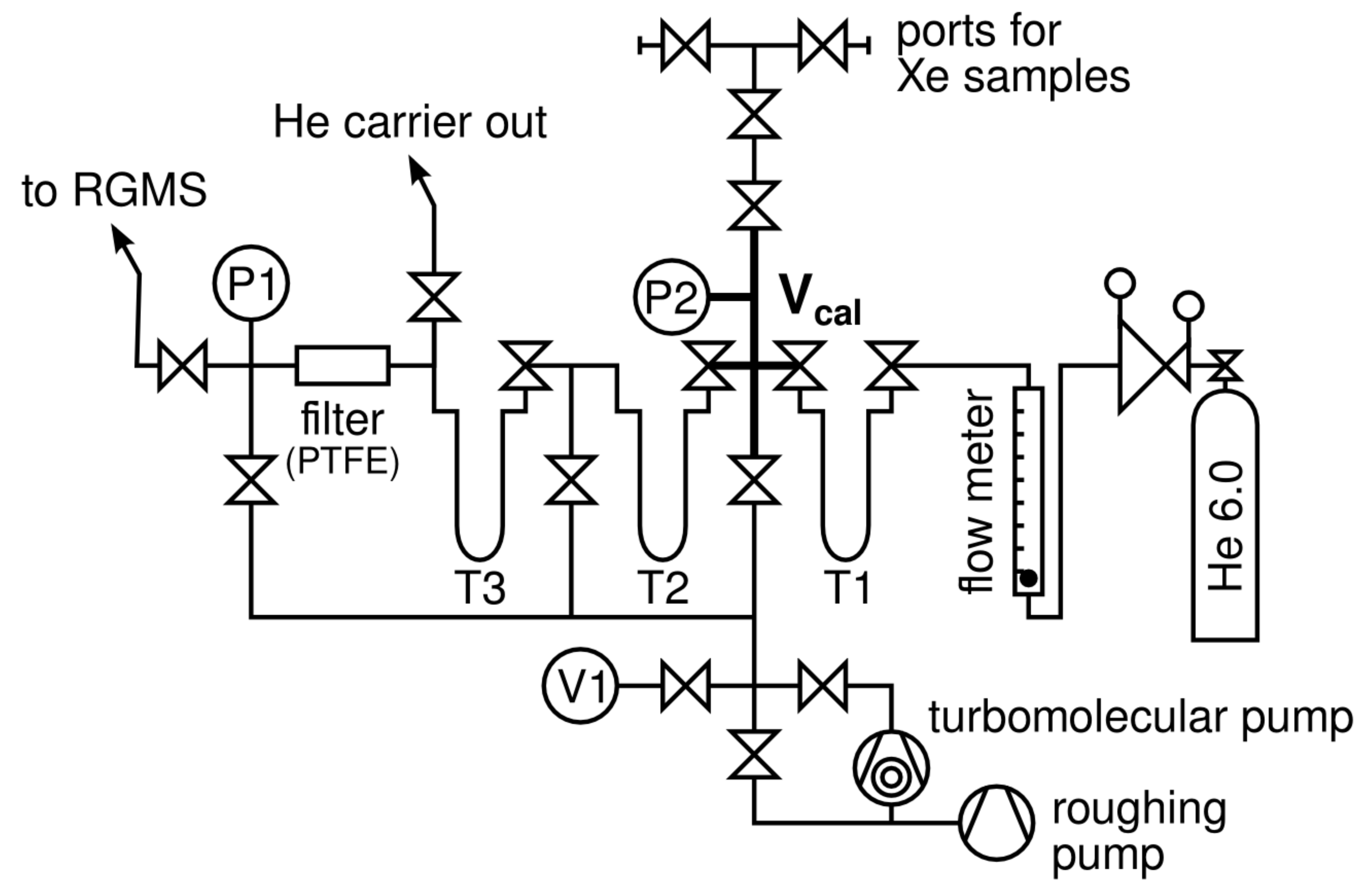}
	\caption{Schematic of the gas chromatography system}
  \label{fig:gchookup}
\end{figure}

\begin{figure}
  \centering
  \includegraphics[width=0.95\columnwidth]{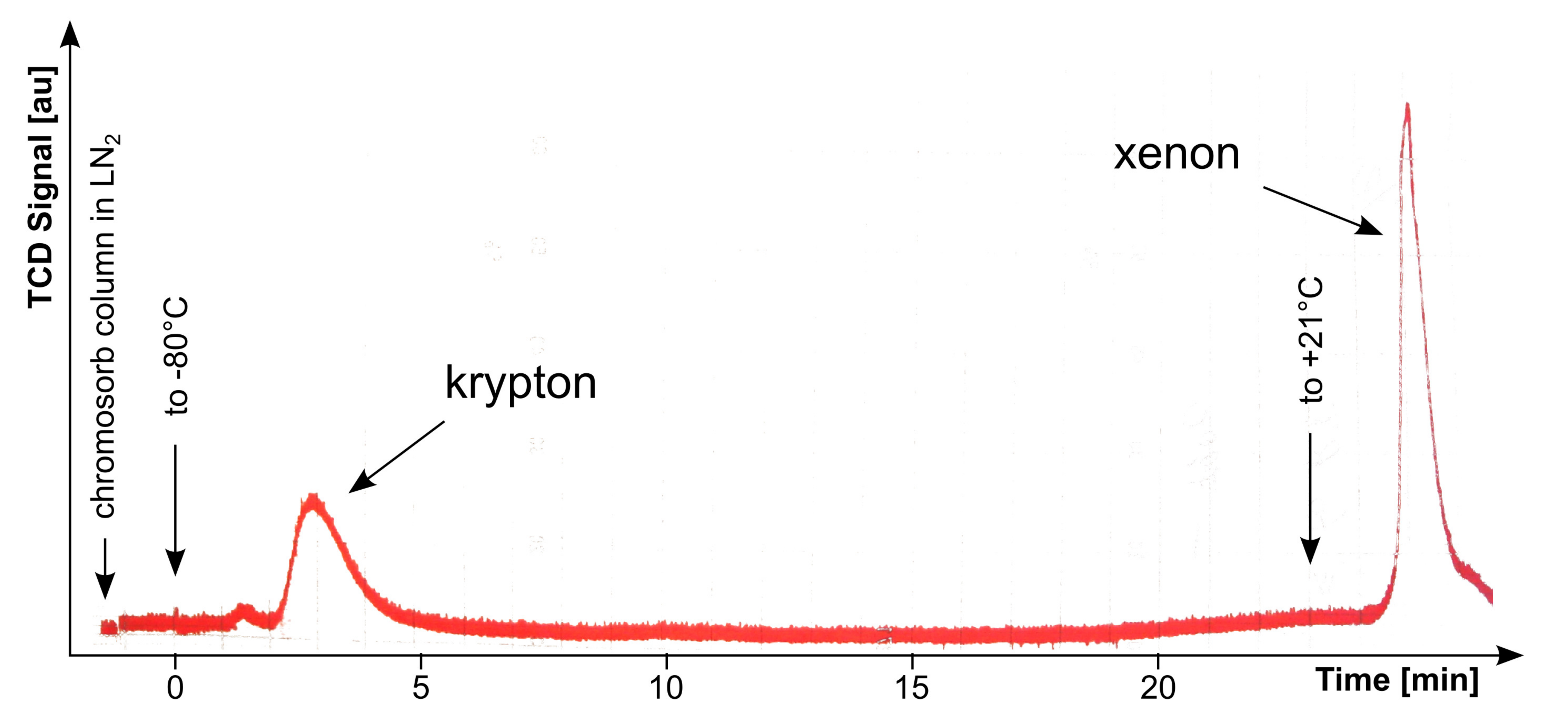}
	\caption{Signal of a thermal conductivity detector  versus time for a
		krypton/xenon separation (\SI{0.5}{ccm} krypton, \SI{1.0}{ccm}
		xenon) via cryogenic gas chromatography using an adsorbent filled packed column
		identical to T2 of the gas chromatographic system}
  \label{fig:krxesep}
\end{figure}

The krypton/xenon separation is performed via gas chromatography in the setup
sketched in Fig.~\ref{fig:gchookup}.
The amount of helium carrier gas used in this process exceeds the size of
the xenon batch by more than a factor of 50. Therefore, the purity
specifications in terms of krypton inside the helium gas are strict. To
reach sub-ppt sensitivity, the krypton concentration in the helium has to
be at or below the ppq level. In this work grade 6.0 helium is purified
using an adsorbent filled packed column T1 (\SI{10}{\mm} inner diameter,
\SI{8.18}{\gram} Carbosieve S-III by Supelco Analytical) immersed in a
liquid nitrogen bath.
Pushed forward by the helium carrier, the gas mixture passes another
adsorbent filled column T2 (\SI{6}{\mm} inner diameter, \SI{0.64}{\gram}
Chromosorb 102 by Johns Manville) immersed in a coolant liquid (ethanol).
Due to differences in the interaction strength of each constituent with the
adsorbent, krypton and xenon can be separated if the flow of the
carrier gas, the pressure gradients and the temperature are properly
adjusted (see Fig.~\ref{fig:krxesep}). The temperature of the ethanol
coolant and the He gas flow for the separation in this work were fixed to
$-\degc{80}$ and 7 standard cubic centimetre per minute (sccm).

The separated krypton is trapped in a third adsorbent filled packed column
T3, identical to T2. By heating up T3 after the chromatographic process the
krypton is released and can be transfered to the mass spectrometer by
cryogenic pumping to the cold finger mounted next to the ion source of the
mass spectrometer.

Apart from the three adsorbent filled traps, the ultra-high vacuum (UHV)
system is constructed completely from stainless steel, and all flanges are
sealed with copper gaskets. The adsorbent filled traps are made from
borosilicate glass with special glass-to-metal seals well suited for the
high purity demands. To avoid dust from the adsorbent entering the mass
spectrometer, a PTFE nano particle filter is installed. The stainless steel
and glass part of the system is bakeable to above \degc{300}, however the
adsorbent materials and the PTFE can only withstand temperatures up to
\degc{150} and \degc{120}, respectively, setting the upper limit on the
temperature during bakeout.  The system provides a turbomolecular pump
backed up by a dry and oil free piston vacuum pump. The latter one is
important to avoid hydrocarbons entering the system and altering the
performance of the adsorbents. All of the valves used in the system are
full-metal, bellow sealed, \SI{3/4}{in.} valves with a leak rate
specification of \SI{<5 e-11}{\mbar\litre\per\second}. A precise
capacitance manometer from Edwards (Barocel 600 series, \perc{0.15}
accuracy, labelled P2 in Fig.~\ref{fig:gchookup}) is installed to determine
the initial batch size of xenon gas. The system is further characterized by
an analog, fully metal sealed pressure gauge (P1) and a combinded
Pirani/cold cathode vacuum gauge (V1) that can be separated from the sample
by a valve.

\subsection{Drawing samples}\label{sec:drawing_samples}

In terms of sampling, two different tasks have to be accomplished: First,
the assay of xenon that must be extracted in situ from immobile experiments
and, second, the assay of xenon in movable storage containers, e.g. xenon
gas cylinders. While the latter type of samples can be shipped to the
laboratory and mounted directly at the device, the sampling of the former
requires the use of an intermediate transport vessel.

\textbf{Sampling from gas cylinders:} The gas chromatographic setup
facilitates two ports (see Fig.~\ref{fig:gchookup}) where gas cylinders can
be connected to either directly or via a pressure regulator. The former has
the advantage of having less volumes potentially contaminated, but is
difficult since the gas cylinders may be filled with pressures up to 65
bar.

\textbf{Sampling with the \xepip{}} Initially developed for a similar
purpose in the framework of the Borexino experiment \cite{Zuzel2004197},
four \SI{1/2}{in.}, fully metal sealed valves were welded together in
series enclosing three volumes of \SI{\sim 10}{\ccm} to be filled with
pressures up to \SI{70}{\bar}. The central volume is shielded by the two
adjacent volumes from ambient air that might penetrate the sample through
tiny leaks. 

The initial preparation of the \xepip{} consists of baking ($\sim
\degc{100}$) and evacuating the three volumes that will hold the samples to
a pressure \SI{<e-7}{\mbar} for several days. Then the sample volumes are
sealed, left under vacuum for several days and finally checked for
remaining krypton background using the mass spectrometer. Once the krypton
background is sufficiently low, the device can be used for taking samples.

The preparation at the site of the xenon extraction consists of mounting the
\xepip{} to a port of the setup that contains the xenon and setting up a vacuum
system. Next, the parts that have been in contact with air, that is between the
setup's port and the first valve of the \xepip{}, are baked and evacuated to a
pressure below \SI{e-7}{\mbar}. To remove possible impurities from the
surfaces and to avoid effects of different impurity compositions in the xenon
gas that might be present close to the port, a small amount of xenon is flushed
from the setup to the vacuum system. Then the valve to the vacuum system is
closed and the samples are drawn by expanding the xenon into the prepared
volumes of the \xepip{}.

\section{\label{sec:calibration}Calibration and data analysis}

A gas standard is available to calibrate the mass spectrometer and to determine
the acceptance of the full gas chromatographic process.  The gas standard
consists of a metal-sealed container with a volume of a few liters that is
filled with the noble gases argon, krypton, and xenon.  Connected to this
container is a second, much smaller volume (\SI{\sim 0.2}{\ccm}) that is
separated from the latter by a fully metal sealed valve. This volume is used to
extract a constant fraction of the gas standard (standard pipette). The amount
of krypton in one standard pipette was calibrated using a sample of
commercially available helium containing a well-known admixture of krypton.
This calibration defines the amount of krypton in our standard pipette to be
\SI{3.9e-11}{\ccm} ($\pm \perc{16.5}$).

\subsection{\label{sec:sem_response}SEM response}

\begin{figure}
  \centering
	\includegraphics[width=0.99\columnwidth]{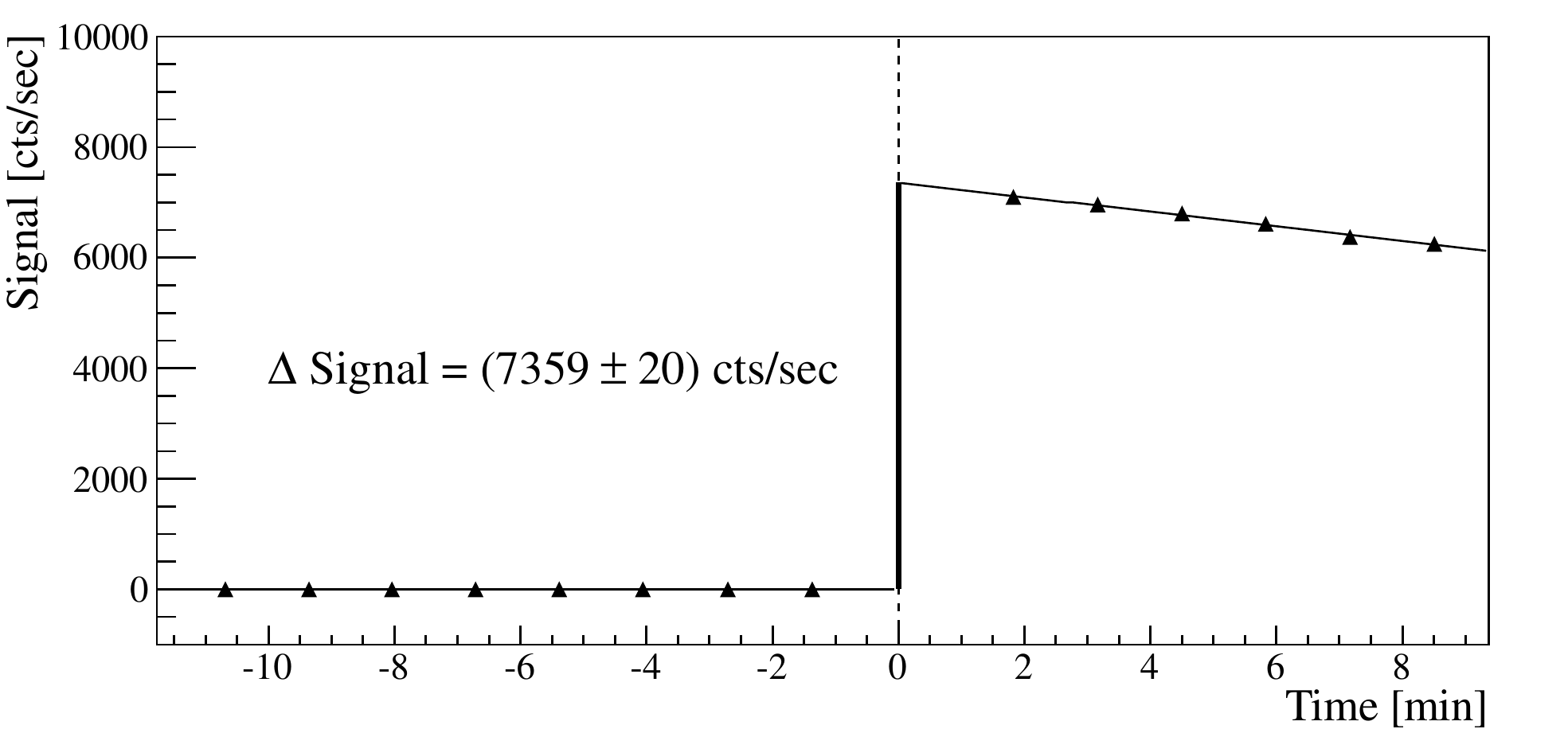}
	\caption{\label{fig:time_fit_kr} Typical time evolution of the krypton
	signal strength from the standard pipette. At time zero
	($\textrm{t}_0=0$) the gas is expanded from the cold finger into the ion
	source and adjacent volumes (vacuum flight tube, detectors)}
\end{figure}

The response of the mass spectrometer to the amount of one standard pipette is
determined at least once before the measurement of a xenon batch. In the
analysis we are interested in the amount of gas initially transfered to the
cold finger, that is at time zero ($\textrm{t}_0$), before other processes
(like implanting accelerated ions into the walls of the mass spectrometer's
vacuum tube) start to change its amount. The time-dependence of the count rate
of the background and signal are fitted and extrapolated to $\textrm{t}_0$
where the size of the pedestal, as seen in Fig.~\ref{fig:time_fit_kr}, is a
measure of the amount of gas from the respective isotope in the batch.

In standard data taking mode a loop over the isotopes \ar{36}, \ar{40},
\kr{84}, \kr{86} and \xe{132} is recorded. The ratios of \ar{40}/\ar{36}
and \kr{84}/\kr{86} provide an intrinsic cross-check since they should
match the natural abundances (mole fractions).  The amount of gas at time
zero is determined for \kr{84} and \kr{86} independently (see
Sect.~\ref{sec:KIL}). From both isotopes the amount of natural krypton is
computed using the respective relative abundance. Both numbers are combined
in the final result.
The gas amount of the initial xenon batch is determined in a calibrated
volume ($\textrm{V}_\textrm{cal}$) equipped with a precise capacitive
pressure gauge (P2, see Fig.~\ref{fig:gchookup}).

\subsection{Krypton acceptance}\label{sec:kr_acceptance}
The
standard pipette is used to determine the acceptance
of the gas chromatographic process for krypton. The calibrated mixture
of krypton and xenon is transferred by cryogenic pumping to the adsorbent
trap T2 (immersed in liquid nitrogen) used
for the krypton/xenon separation. Then the full chromatographic procedure is performed --
identical to the standard procedure for a normal xenon batch -- and the
spectrometer's response to the krypton isotopes is compared
to its response without the chromatographic separation process. This calibration
procedure yields a krypton acceptance of $\epsilon_a = (0.97 \pm 0.02)$.

\sloppy{
The procedure described above employs microscopic krypton and xenon gas
samples on the order of a few \SI{e-12}{\ccm}, while the standard procedure
foresees xenon batches of up to several \si{\ccm} containing only traces of
krypton. When combining \SI{3.9e-11}{\ccm} of \kr{nat} from the standard
pipette with approximately \SI{1.2}{\ccm} of xenon from sample XE100-3
(\SI{1.8}{\ppt} intrinsic krypton, see Sect.~\ref{sec:results}), we
measure \SI{3.2(7)e-11}{\ccm} \kr{nat}. Reevaluating the acceptance for
krypton $\epsilon_a$, we find $\epsilon_a = (0.79 \pm 0.18)$ in agreement
with above result. The larger uncertainty in this value is dominated by the
uncertainty in the intrinsic krypton level of the bulk xenon and the sample
standard deviation of a single measurement (see
Sect.~\ref{sec:error_budget}). The weighted average of both, $\epsilon_a
= (0.97 \pm 0.02)$, will be used throughout this work.}

\subsection{\label{sec:KIL}Computation of the krypton level}
The krypton level of a given batch is the ratio of the amount of
krypton gas measured by the mass spectrometer and the initial batch size
determined by its pressure in the calibrated volume. We can
write:

\begin{equation}
  \frac{\kr{nat}}{\xe{nat}} = \frac{\left\langle\left( \kr{i}
	- B_i  \right) \cdot f_m(i)^{-1} \right\rangle_i \cdot \epsilon_a^{-1}}{\xe{nat}} \, ,
  \label{equ:comp_il}
\end{equation}
where $i$ represents the individual isotope, $f_m(i)$ is the mole fraction of
the respective isotope, $\langle\dots\rangle_i$ denotes the average over
the isotopes $i$  and $\epsilon_a$ is the
acceptance of the full chromatographic process for krypton.  
$\kr{i}$ is the amount of gas of the krypton isotope
$i$ as measured with the spectrometer. It is computed from the abundance of the krypton
isotope $i$ in the standard pipette ($\textit{std}_{i}$) and from the ratio
of the pedestals of the batch ($s_i^\text{batch}$) and the standard pipette
($s^\text{std}_{i}$):
\begin{equation}
  \kr{i} = \frac{s_i^\text{batch}}{s^\text{std}_{i}} \cdot
  \textit{std}_{i}\, .
  \label{equ:comp_kr_i}
\end{equation}
Finally, $B_i$ is the procedure blank of the full process for the isotope
$i$, as will be explained in Sect.~\ref{sec:procedure_blank}, and can be
written similar to (\ref{equ:comp_kr_i}) substituting $s_i^\text{batch}$ by
$s_i^{B}$.  $^{nat}$Xe can be computed, if the pressure $p$ and the
temperature $T$ in the calibrated volume $\text{V}_\text{cal}$ are known.

\subsection{Uncertainty budget estimation}\label{sec:error_budget}
Systematic effects common to all measurements are the
uncertainty of the standard pipette, the uncertainty in determining the
initial xenon batch size and the uncertainty of the chromatographic
acceptance for krypton ($\Delta \epsilon_a$). Summing them quadratically, we
end up with a relative uncertainty of \perc{17.0} dominated by the
uncertainty of the standard pipette.

Statistical fluctuations of the number of ions detected by the SEM,
variations in performing the chromatographic separation and variations in
the electrical fields of the ion optics due to high voltage drifts
individually affect single measurements. The sum of these effects is
estimated (for each sample individually) from the fluctuations of the
measurement of several batches around their mean. The resulting uncertainty
estimator is added quadratically to the aforementioned systematic
uncertainties. In those cases where only a single batch was measured, we
estimate its uncertainty by averaging the sample standard deviations of all
measurements so far done with this setup.  We find \perc{13.5} to be a
single measurement's standard deviation.  Combining both uncertainties by
summing them quadratically we find the total relative uncertainty of a
single batch to be \perc{21.7}. If more than one single batch is measured,
the relative uncertainty decreases being limited to the
systematic uncertainty of \perc{17}.

\subsection{Sensitivity  -- Procedure Blank and Detection
Limit}\label{sec:procedure_blank}

The sensitivity is limited by the traces of krypton which are collected during
the full process and increase the signal in the mass spectrometer. Sources
for this type of krypton are potential microscopic leaks allowing external krypton to
enter the system, outgassing of the surfaces involved in the measurement
and krypton introduced by the helium carrier gas.
The amount of these krypton traces is accessed by performing the full
procedure, identical to a normal measurement, but without a xenon batch
introduced (procedure blank). A value of \SI{1.00(4)e-12}{\ccm} is found
averaging several measurements done between the presented xenon samples of
Sect.~\ref{sec:results}. 
After a recent system upgrade (refurbished high voltage control of ion
optics and identification and removal of tiny air leak in the sample
preparation part) the current procedure blank was measured to be
\SI{0.081(4)e-12}{\ccm}.  This is a reduction by more than a factor ten
with respect to the value mentioned above. The first xenon sample measured
at this significantly reduced background level is in excellent agreement
with earlier results at the higher background level (see
Sect.~\ref{sec:results}).  The systematic uncertainty of the absolute
calibration is neglected in the estimate of the uncertainty of the
procedure blank. This is justified as measurements are corrected for the
procedure blank before being converted from a count rate to an amount of
gas using the absolute calibration factor.

The decision threshold (DT) and detection limit (DL) are computed following
\cite{ISO11929:2010} and result in \SI{0.007e-12}{\ccm} and
\SI{0.015e-12}{\ccm} (\kr{nat}), respectively, assuming a
\perc{21.7} uncertainty of the outcome of a single measurement. The maximal
size of the xenon batch that can be processed by the gas chromatographic
separation determines the performance of the system. Up to now the largest
xenon batch was \SI{1.9}{\ccm} and we find at \perc{95} confidence
level:
\begin{equation}
  \text{DL} = \frac{(\kr{nat})_\text{min}}{(\xe{nat})_\text{max}} =
	\SI{8}{\ppq}
  \label{equ:detection_limit}
\end{equation}
Note that likely larger xenon batches can be processed by the chromatographic
process. The upper threshold for acceptable batch sizes was not determined
yet. Consequently, the detection limit of currently \SI{8}{\ppq} may still be
lowered in the future.

\section{Results}\label{sec:results}

Table~\ref{tab:rgms_results_details} lists the results of the individual
measurements together with the amount of processed xenon gas. In total,
measurements of seven different samples are presented. Two of them (CYL-1
and CYL-2) are xenon gas cylinders with krypton concentrations at the low
ppb  and high ppt level, i.e. sufficiently abundant to be measured with
more common devices for cross checks. The remaining five samples were drawn
in the context of the XENON100 experiment \cite{Aprile:2011dd}. 

\begin{table*}
	\centering
	\caption{\label{tab:rgms_results_details}Details on the individual
	measurements presented in this work} 
	\begin{tabular}{lll}
	\hline  
	Name & \xe{nat} $[10^{-3}\units{cm}^3]$ & Kr/Xe $[\units{ppt}]$ \\ 
	\hline
  CYL-1  & $452 \pm 13$ & $(7.83 \pm 0.24^\text{stat} \pm 1.33^\text{sys})\times 10^3$ \\
      & $96 \pm 4$ & $(7.37 \pm 0.27^\text{stat} \pm 1.25^\text{sys})\times 10^3$ \\
      & $1710 \pm 50$ & $(6.59 \pm 0.20^\text{stat} \pm 1.12^\text{sys})\times 10^3$ \\
      & $158 \pm 5$ & $(7.72 \pm 0.25^\text{stat} \pm 1.31^\text{sys})\times 10^3$ \\
		  &  & $\mathbf{(7.3 \pm 1.3)\times 10^3}$ \\ 
	CYL-2 & $1240 \pm 40$ & $294 \pm 9^\text{stat} \pm 50^\text{sys}$ \\
      & $1810 \pm 50$ & $310 \pm 9^\text{stat} \pm 53^\text{sys}$ \\
      & $108 \pm 4$ & $277 \pm 10^\text{stat} \pm 47^\text{sys}$ \\
		  &  & $\mathbf{290 \pm 50}$ \\ 
  XE100-1  & $603 \pm 17$ & $353 \pm 11^\text{stat} \pm 60^\text{sys}$ \\
       & $586 \pm 17$ & $359 \pm 10^\text{stat} \pm 61^\text{sys}$ \\
       & $669 \pm 19$ & $369 \pm 11^\text{stat} \pm 63^\text{sys}$ \\
       & $599 \pm 17$ & $279 \pm 8^\text{stat} \pm 47^\text{sys}$ \\
	     &  & $\mathbf{340 \pm 60}$ \\ 
  XE100-2  & $832 \pm 24$ & $14.5 \pm 0.5^\text{stat} \pm 2.5^\text{sys}$ \\
        & $191 \pm 6$ & $13.1 \pm 1.4^\text{stat} \pm 2.2^\text{sys}$ \\
        & $615 \pm 18$ & $13.9 \pm 0.6^\text{stat} \pm 2.4^\text{sys}$ \\
        & $692 \pm 20$ & $13.5 \pm 0.6^\text{stat} \pm 2.3^\text{sys}$ \\
		  &  & $\mathbf{14.0 \pm 2.0}$ \\ 
  XE100-3  & $398 \pm 12$ & $1.80 \pm 0.14^\text{stat} \pm 0.31^\text{sys}$ \\
	  &  & $\mathbf{1.8 \pm 0.4}$ \\ 
  XE100-4  & $700 \pm 20$ & $0.97 \pm 0.09^\text{stat} \pm 0.16^\text{sys}$ \\
      & $170 \pm 5$ & $1.0 \pm 0.4^\text{stat} \pm 0.2^\text{sys}$ \\
	 	  &  & $\mathbf{0.97 \pm 0.19}$ \\ 
  XE100-5   & $262 \pm 8$ & $0.71 \pm 0.18^\text{stat} \pm 0.12^\text{sys}$ \\
      & $1950 \pm 60$ & $0.96 \pm 0.03^\text{stat} \pm 0.16^\text{sys}$ \\
		  &  & $\mathbf{0.95 \pm 0.16}$ \\ 
	\hline
	\end{tabular}
	\caption*{The column Kr/Xe lists the
	krypton concentration specifying statistical and systematic uncertainties. The
	final numbers combining multiple measurements (if available) are given in bold face.}
\end{table*}

\textbf{CYL-1}
Sample drawn from a carbon steel gas cylinder containing high-purity xenon
depleted in the isotope \xe{136} from the supplier Iceblick.  Combining the
four measurements and adding the systematic uncertainty as described in
Sect.~\ref{sec:error_budget}, we find a krypton level of
\SI{7.3(13)}{\ppb}. With this level in the low ppb regime a completely
independent measurement using a customized gas chromatograph (Trace GC
Ultra from Thermo Scientific) employing a pulsed helium discharge
photo-ionization detector (He-PDPI) was possible. This measurement
(performed also at MPIK) yielded a krypton concentration of
\SI{6.3(13)}{\ppb} in agreement with the mass spectroscopic result.

\textbf{CYL-2} 
Sample drawn from an aluminum gas cylinder containing high-purity xenon
from the supplier Air Liquide. The xenon was purchased by our colleagues
from the University of M{\"u}nster and measured also by their device
combining a cold-trap with a quadrupole mass-spectrometer. The three
measurements done with our system combine to a final value of
\SI{0.29(5)}{\ppb}. This is again in good agreement with the completely
independent analysis from our colleagues that found a krypton concentration
of \SI{0.33(20)}{ppb} in this very same gas sample
\cite{2013JInst..8P2011B}.

\textbf{XE100-1}
Sample drawn from the XENON100 detector \cite{Aprile:2011dd} during dark
matter data taking of the first XENON100 science run \cite{Aprile:2011hi}.
In total there are four measurements with a combined result of
\SI{340(60)}{\ppt}. In \cite{Aprile:2011hi} a krypton/xenon level of
\SI{700(100)}{\ppt} is given based on a spectral fit to \kr{85}. This result
was obtained assuming a $\kr{85}/\kr{\text{nat}}$ ratio of $1\times
10^{-11}$. Using the better motivated number of $2\times 10^{-11}$
\cite{2003GeoRL..30tHLS4D,Aprile:2012nq}, the modified value of
\SI{350(50)}{\ppt} agrees with the result obtained in this work.

\textbf{XE100-2}
During the data taking of the second XENON100 science run
\cite{Aprile:2012nq}, a sample was drawn again from the XENON100 detector.
In total four measurements were performed resulting in \SI{14(2)}{\ppt}.
Note the different value of \SI{19(4)}{\ppt} reported in
\cite{Aprile:2012nq}, that was evaluated in advance of the precise
calibration of our system reported in this work. The correct number of
\SI{14(2)}{\ppt} is in agreement with the value of \SI{18(8)}{\ppt} obtained
from counting the number of delayed $\beta - \gamma$ coincidences
associated with the \kr{85} beta decay \cite{Aprile:2012nq}. Here the
$\kr{85}/\kr{\text{nat}}$ ratio of $2\times10^{-11}$ is used.

\textbf{XE100-3} 
Sample drawn from the XENON100 gas system during a maintenance
period after the second science run. Taking into account the systematic
uncertainty, we measure a value of \SI{1.8(4)}{\ppt} for the
one measurement done of this sample.

\textbf{XE100-4}
Sample drawn from the output of the XENON100 cryogenic distillation column
(similar to the one described in \cite{2009APh....31..290A}) optimized to
remove krypton from xenon gas.  Combing the results of the two measurements
yields a value of \SI{0.97(19)}{\ppt} for this sample. While the
first xenon batch had a size of about \SI{0.7}{\ccm} giving a signal
well above the detection limit, the second xenon batch was smaller by a
factor of four.  Still the central value is in very good agreement with the
first measurement proving the linearity of the spectrometer for such small
signals. Comparing the size of this second xenon batch with the second one
from XE100-5 (Table~\ref{tab:rgms_results_details}) we see that the amount
of xenon can be increased by a factor of ten giving sensitivity to krypton
concentrations below \SI{100}{\ppq} as suggested by the detection limit
(Eq.~\ref{equ:detection_limit}).

\textbf{XE100-5}
The last two samples reported here were again drawn from the XENON100
detector. A single measurement of the first sample resulted in
\SI{0.71(24)}{\ppt}. At the significantly improved background conditions of
our system the second sample was measured to contain \SI{0.96(22)}{\ppt}.
Combining both results we find a krypton concentration in the XENON100
detector of \SI{0.95(16)}{\ppt}. To our knowledge, this is the purest xenon
target ever employed in a LXe particle detector.

\section{\label{sec:summary}Summary}
We have presented our new approach of assaying krypton concentrations in
xenon at the ppq level using a gas chromatographic krypton/xenon separation
followed by a mass spectroscopic krypton measurement. We explained the data
analysis and absolute calibration of our device and discussed in detail its
uncertainties. By comparing our assay results of two xenon gas samples at
the low ppb and high ppt level with results obtained by two independent
devices, we confirmed the thorough understanding of the absolute
calibration and linearity of the instrument.  We showed that with the
current setup the system's detection limit is 8 ppq which may be further
improved by enlarging the xenon sample size.

\paragraph{Acknowledgements}
{\small
We are grateful to the XENON collaboration for their support in drawing
xenon samples and providing the purest xenon gas. We greatly acknowledge
the help of Jens Hopp, Hans Richter, Dominik Stolzenburg and Grzegorz Zuzel
for their support with the sector field mass spectrometer and thank Teresa
Marrod\'{a}n Undagoitia for her support during the final stage of this
work.}  

\bibliographystyle{apsrev4-1}

%

\end{document}